\documentclass[prl,twocolumn,letter, showpacs, superscriptaddress]{revtex4}
\usepackage{amssymb}
\usepackage{amsmath}
\usepackage{bbm}
\usepackage{graphicx}

\begin{document}

\author{Philipp Werner}
\affiliation{Columbia University, 538 West, 120th Street, New York, NY 10027, USA}
\author{Emanuel Gull}
\author{Matthias Troyer}
\affiliation{Theoretische Physik, ETH Zurich, 8093 Zurich, Switzerland}
\author{Andrew J. Millis}
\affiliation{Columbia University, 538 West, 120th Street, New York, NY 10027, USA}
\title{Spin freezing transition and non-Fermi-liquid self-energy in a 3-orbital model}
\date{\today}

\hyphenation{}

\begin{abstract}
A single-site dynamical mean field study of a three band model  with the rotationally invariant interactions appropriate to the $t_{2g}$ levels of a transition metal oxide  reveals a quantum phase transition between a paramagnetic metallic phase and an incoherent metallic phase with frozen moments. The Mott transitions occurring at electron densities $n=2,3$ per site take place inside the frozen moment phase.  The critical line separating the two phases is characterized by a self energy with the frequency dependence $\Sigma(\omega)\sim \sqrt{\omega}$ and  a broad quantum critical regime. 
The findings are discussed in the context of the power law observed in the optical conductivity  of SrRuO$_3$.
\end{abstract}


\maketitle

The `Mott' metal-insulator transition plays a central role in the modern conception of strongly correlated materials \cite{Imada98,Nagaosascience}. Much of our understanding of this transition comes from studies of the one-band Hubbard model. Here,  the transition is generically masked by antiferromagnetism, but if this is suppressed (physically, by introducing lattice frustration or mathematically, by examining an appropriately restricted class of theories such as the paramagnetic-phase single site dynamical mean field approximation \cite{Georges96}) a transition from a paramagnetic metal to a paramagnetic insulator is revealed. The properties of the paramagnetic metal phase near the transition play a central role in our understanding of the physics of correlated electron compounds. 

While one band models are relevant to many materials including the high temperature superconductors and  some organic compounds, many systems of interest involve multiple correlated orbitals for which the physics is richer and less fully understood.  Multiorbital models have been studied in Refs.~\cite{Florens02,Ono03,Keller04,Biermann05,Inaba05,Koga05,Lombardo05}.  New physics related to the appearance of magnetic moments has been considered in the context of the  orbitally selective Mott transition which may occur if the orbital degeneracy is lifted \cite{Anisimov02,Liebsch05,Biermann05b,Werner07Crystal,Costi07},  but for orbitally degenerate models it seems  accepted that the essential concepts 
of a paramagnetic metal to paramagnetic insulator transition and a strongly correlated paramagnetic metal phase can be carried over from studies of the one-band situation. 

In this paper we show that this assumption is not correct. We use the single-site dynamical mean field approximation to demonstrate the existence of a quantum phase transition between a paramagnetic Fermi liquid and an incoherent metallic phase characterized by frozen local moments (a spin-spin correlation function which does not decay to zero at long times). We show that for densities per site $n=2,3$ the Mott transition occurs within or at the boundary of the frozen moment phase. As Costi and Liebsch have noted in the context of an orbitally selective Mott system,  the presence of  frozen moments may be expected to influence the Mott transition \cite{Costi07}. The new phase appears for multiple orbitals, a different number of electrons than orbitals 
and a rotationally invariant on-site exchange $U/3>J>0$.  While this situation is relevant to many compounds, it has has only recently become amenable to systematic study following the development of   improved exact diagonalization \cite{Perroni07} and 
efficient Monte Carlo methods \cite{Werner05,Werner06}. 

The transition into the frozen moment phase produces a non-Fermi liquid self energy which varies as $\sqrt{\omega}$, and the quantum critical regime of the phase transition is sufficiently large that the non-Fermi liquid behavior is observable over a wide range of temperature, interaction strength, spin polarization and carrier concentration.  The non-Fermi-liquid self energy we find bears an intriguing resemblance to the self energy inferred from optical conductivity measurements on SrRuO$_3$ \cite{Kostic98,Dodge02,Lee02}.

We study a model inspired by titanate, vanadate and ruthenate transition metal oxide compounds, in which the relevant electrons reside in  $t_{2g}$-symmetry $d$-orbitals. In the commonly occurring pseudocubic structure the $t_{2g}$ levels are three-fold degenerate, and the formal valence is such that the levels contain $1$ electron (Ti compounds),  $2$ electrons (some V compounds) or $4$ electrons / $2$ holes (Ru compounds).  The same physics occurs for the two-orbital models relevant to the nickelates and other $e_g$-based systems. The electronic Hamiltonian  is $H=H_\text{band}+\sum_iH_\text{loc}^i,$ with   $H_\text{band}=\sum_{\alpha,p} \varepsilon^\alpha_p d^{\alpha\dagger}_{p,\sigma}d^\alpha_{p,\sigma} $ and the local Hamiltonian on each site $i$ given by 
\begin{eqnarray}
&&H_\text{loc}=-\sum_{\alpha, \sigma}\mu n_{\alpha,\sigma}+\sum_{\alpha} U n_{\alpha,\uparrow} n_{\alpha,\downarrow}\label{H_loc}\\
&&\hspace{2mm}+\sum_{\alpha>\beta,\sigma} U' n_{\alpha,\sigma} n_{\beta,-\sigma} +  (U'-J) n_{\alpha,\sigma}n_{\beta,\sigma}\nonumber\\
&&\hspace{2mm}-\sum_{\alpha\ne\beta}J(\psi^\dagger_{\alpha,\downarrow}\psi^\dagger_{\beta,\uparrow}\psi_{\beta,\downarrow}\psi_{\alpha,\uparrow}
+ \psi^\dagger_{\beta,\uparrow}\psi^\dagger_{\beta,\downarrow}\psi_{\alpha,\uparrow}\psi_{\alpha,\downarrow} + h.c.).\nonumber
\end{eqnarray}
Here, $\alpha=1,2,3$ is the orbital index, $\sigma=\uparrow, \downarrow$ the spin index, $\mu$ is the chemical potential,  $U$ ($U'$) the intra-orbital (inter-orbital) Coulomb interaction, and $J$ the coefficient of the Hund coupling. We adopt the conventional choice of parameters,  $U'=U-2J$.

To study the model we use the single-site dynamical mean field approximation \cite{Georges96} which ignores the momentum dependence of the self-energy and reduces the original lattice problem to the self-consistent solution of a quantum impurity model given by the Hamiltonian $H_\text{QI} = H_\text{loc}+H_\text{hyb}+H_\text{bath}$ with $H_\text{hyb}$ describing the exchange of electrons between impurity and bath. Our data were computed for a semi-circular density of states with band-width $4t$ (so the model is particle-hole symmetric about the density $n=3$), using the continuous time QMC solver of Ref. \cite{Werner05, Werner06}. We investigate the electron self energy $\Sigma(\omega)$ and the imaginary-time impurity-model spin-spin and orbital-orbital correlators $\langle {\cal O}(\tau){\cal O}(0) \rangle$ with ${\cal O}$ representing either the electron spin density $S_z=\frac{1}{3}\sum_\alpha \frac{1}{2}(d^\dagger_{\alpha,\uparrow} d_{\alpha,\uparrow}-d^\dagger_{\alpha,\downarrow}d_{\alpha,\downarrow})$ or the orbital density
${\hat n}_{\alpha}=\sum_\sigma d^\dagger_{\alpha,\sigma}d_{\alpha,\sigma}$.  Attention is restricted to solutions which preserve spin and orbital rotational symmetry at the level of the impurity model.

\begin{figure}[t]
\begin{center}
\includegraphics[angle=-90, width=0.9\columnwidth]{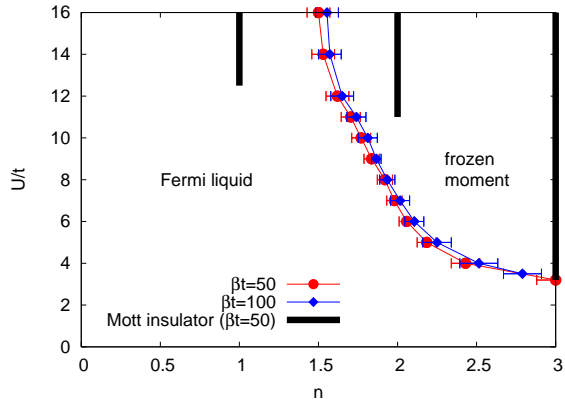}
\caption{Phase diagram for $J/U=1/6$ and $\beta t =50$, $100$ in the space of density $n$ and interaction strength $U$.  The light line with circles or diamonds indicates a phase transition between a Fermi liquid metal and  a ``frozen-moment" metal. The black lines mark the regions of Mott insulating behavior.}
\label{phasediagram}
\end{center}
\end{figure}

Figure \ref{phasediagram} presents our calculated phase diagram in the space of density $n$ and interaction strength $U$ for the ratio $J/U=1/6$. The Mott insulating  phases of the model are shown as heavy solid lines.  The light line with circles or diamonds is our new result: a phase boundary separating a small $n$ small $U$ Fermi liquid phase from a frozen moment phase at larger $n$ and larger $U$. Other values of $0<J/U<1/3$ give  similar results. For $J=0$ the new phase does not exist while for $J>U/3$ the term $U'-J=U-3J$ changes sign and the physics of the model becomes different.  

\begin{figure}[t]
\begin{center}
\includegraphics[angle=-90, width=0.9\columnwidth]{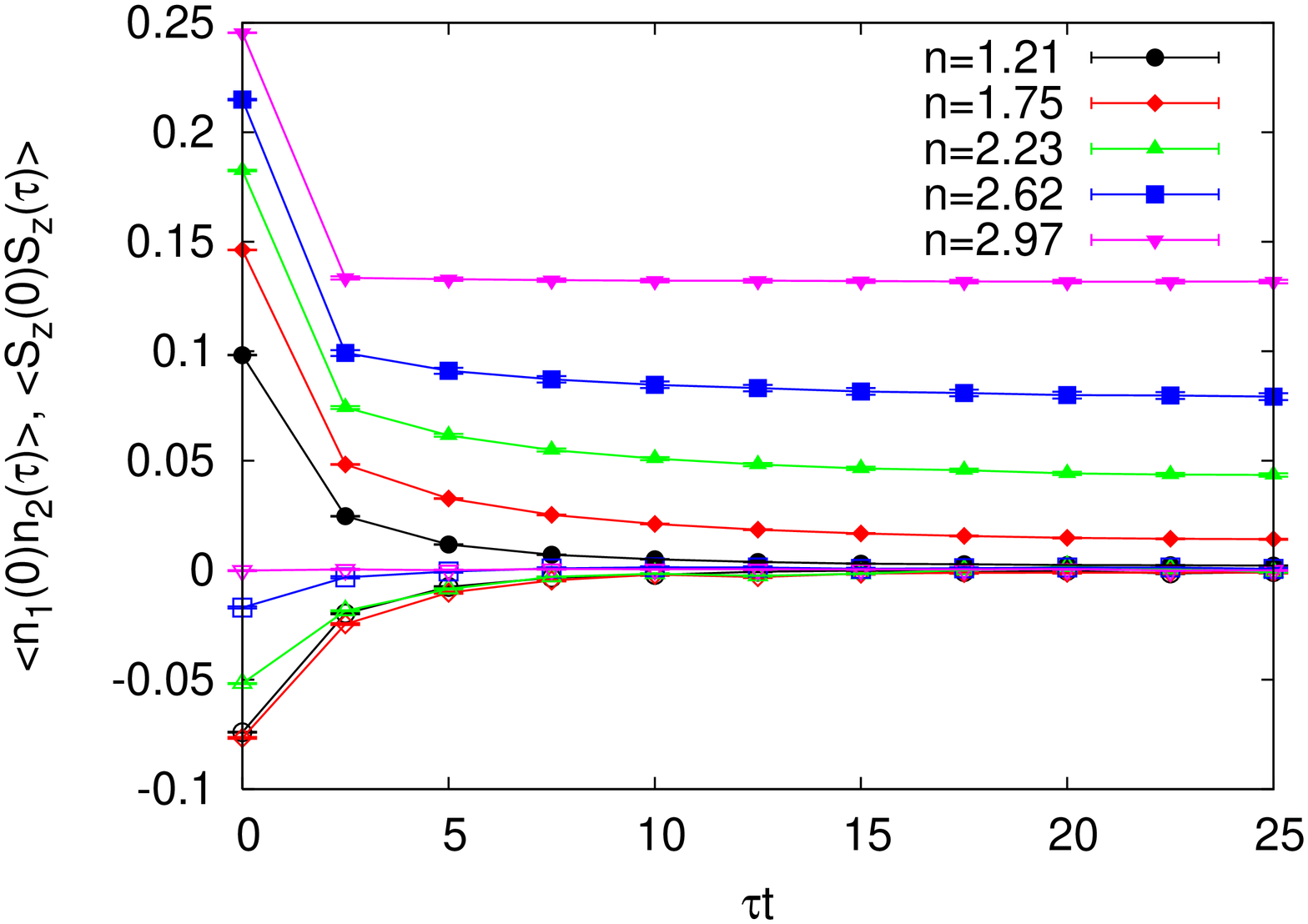}
\includegraphics[angle=-90, width=0.85\columnwidth]{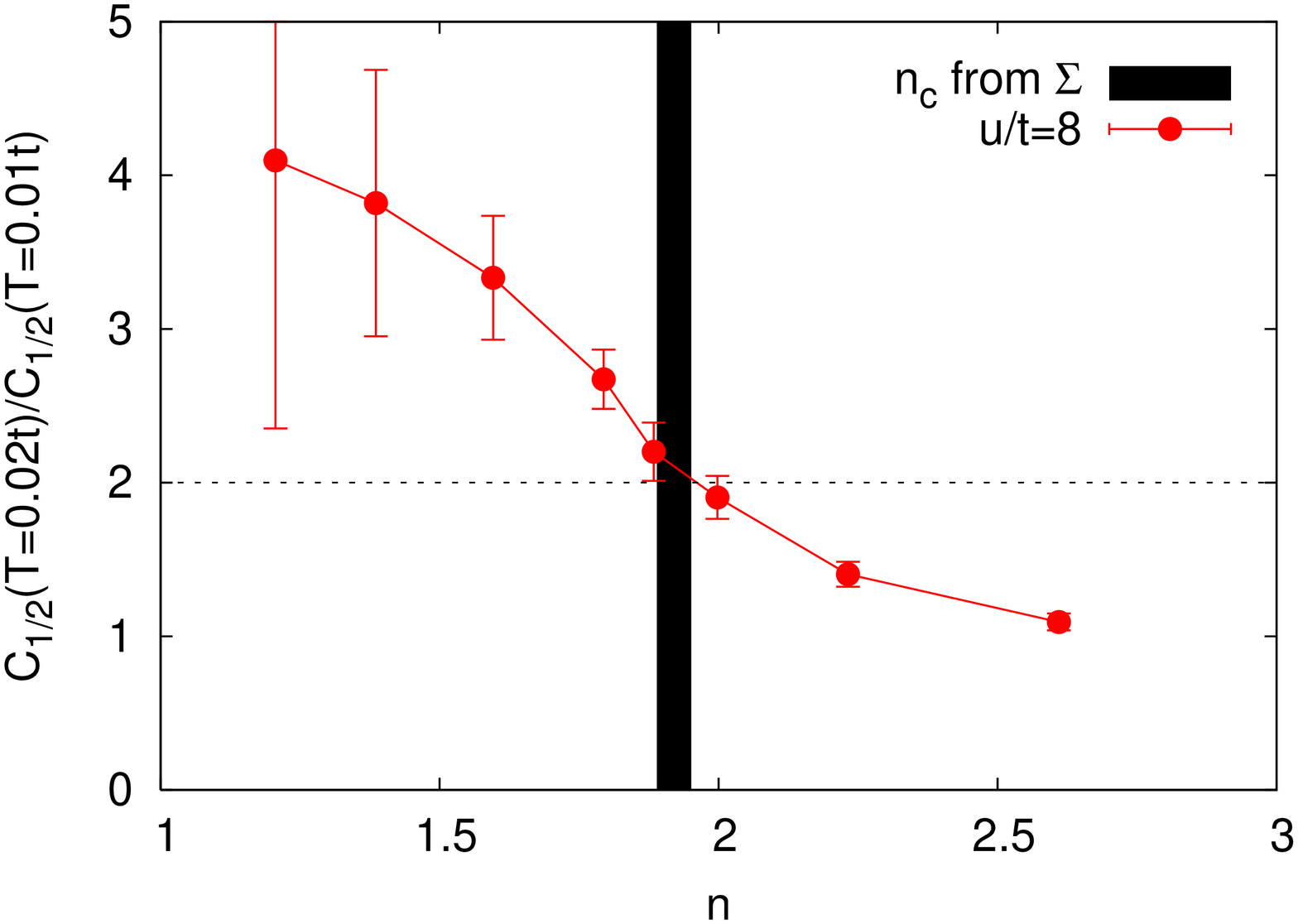}
\caption{Upper panel: imaginary time dependence of the spin-spin correlation function $\langle S_z(0)S_z(\tau)\rangle$ (positive correlation function, full symbols) and orbital correlation function $\langle n_1(0)n_2(\tau)\rangle$ (negative correlation function, open symbols) for $U=8t$ and carrier concentrations $n$ indicated. 
Lower panel: variation with doping of the temperature dependence of the spin-spin correlation at  $\tau=\beta/2$. The error bars are large at smaller $n$ because the midpoint spin correlator is very small. The black line indicates the $n$-value of the phase transition deduced from an analysis of the self energy. 
}
\label{szsz_ninj_doping}
\end{center}
\end{figure}

We may define the phase boundary using the impurity model spin-spin correlation function $C_{SS}(\tau)=\langle S_z(\tau)S_z(0)\rangle$, shown in the upper panel of  Fig.~\ref{szsz_ninj_doping} for $U/t=8$ and several values of $n$.  In a Fermi liquid at low temperature $T$,  $C_{SS}(\tau)\sim (T/\sin(\pi \tau T))^2$  for imaginary times $\tau$ sufficiently far from either $\tau=0$ or $\tau=1/T$. Our results are consistent with this form in the Fermi liquid phase, but in the non-Fermi-liquid phase $C_{SS}$ is seen to approach a constant at long times indicating the presence of frozen moments. We also plot in Fig.~\ref{szsz_ninj_doping} the corresponding orbital correlation function, which is seen to decay rapidly with time on both sides of the phase transition.  For a more quantitative analysis we studied the temperature dependence of $C_{1/2}\equiv C_{SS}(\tau=\frac{1}{2T})$. In a Fermi liquid, $C_{1/2}\sim T^2$ while in the frozen moment phase $C_{1/2}$ becomes temperature independent at sufficiently low $T$.  Within our numerical accuracy, we find that at the transition point, $C_{SS}(\tau=\frac{1}{2T}) \sim T$. The lower panel of Fig.~\ref{szsz_ninj_doping} shows how the ratio $C_{1/2}(T=0.02t)/C_{1/2}(T=0.01t)$ changes from the value $4$ expected in the Fermi liquid phase to the value $1$ expected in the frozen moment phase.  
\begin{figure}[t]
\begin{center}
\includegraphics[angle=-90, width=0.9\columnwidth]{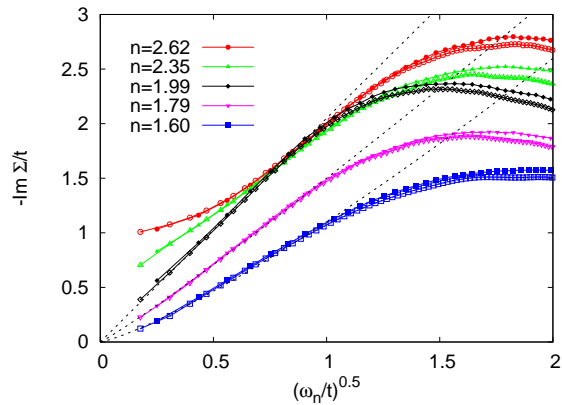}
\caption{Doping dependence of the imaginary part of the Matsubara axis self energy for $U/t=8$, $\beta t=50$ (full symbols) and 100 (open symbols) at dopings indicated. The dashed lines are proportional to $(\omega_n/t)^\alpha$, $\alpha=0.5$, 0.54, 0.62 (from top to bottom).
}
\label{qp_doping}
\end{center}
\end{figure}

\begin{figure}[t]
\begin{center}
\includegraphics[angle=-90, width=0.9\columnwidth]{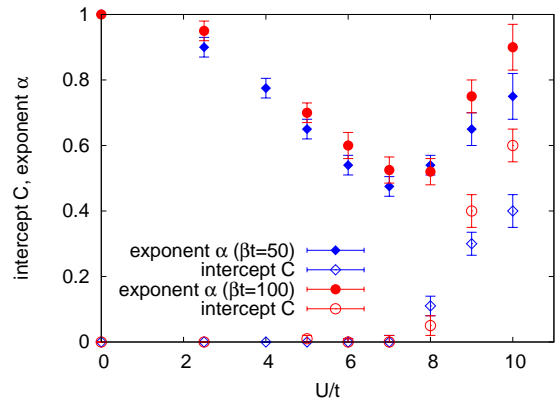}
\caption{
Results of a fit of the computed Matsubara-axis self energy to the scaling form $-\text{Im} \Sigma/t=C+A(\omega_n/t)^\alpha$ at temperatures $T=t/50$ (diamonds; blue on-line) and $T=t/100$ (circles, red on-line) at density $n=2$ and interaction strengths indicated. The plot illustrates the wide quantum critical regime of the spin freezing transition.
}
\label{self}
\end{center}
\end{figure}

The phase transition has consequences for the electron self energy $\Sigma(\omega)$. In a Fermi liquid at low $T$  the imaginary part of the  real axis scattering rate $\Sigma^{''}(\omega)\sim \max (\omega^2,T^2)$ so the imaginary part of the Matsubara axis self energy $\Sigma(i\omega_n) \sim i\omega_n$ at small $\omega_n$. Frozen moments may be expected to scatter electrons so that the real axis self energy is $\Sigma^{''}(\omega)\sim \Gamma$, implying on the Matsubara axis $\text{Im} \Sigma (i\omega_n\rightarrow 0)=i\Gamma \text{sgn} (\omega_n)$. At the critical point we expect a power law behavior $\Sigma^{''}(\omega)\sim \omega^\alpha$; if $\alpha<1$ then $\ \Sigma (i\omega_n\rightarrow 0)\sim (i\omega_n)^\alpha$. Figure~\ref{qp_doping} shows that the imaginary part of our computed Matsubara-axis self energy is consistent with this behavior, vanishing linearly in the Fermi liquid phase, exhibiting an intercept  in the frozen moment phase and an exponent $\alpha \approx 1/2$ at $n_c$.  The behavior of the self energy in the region not too far from the transition is described by a  quantum critical crossover function. Our resolution is not sufficient to identify this function precisely. We have fit the self energy to the form $\text{Im} \Sigma (\omega_n)/t=C+A(\omega_n/t)^\alpha$, recognizing that the varying value of $\alpha$ represents the different regimes of the crossover function. In Fig.~\ref{self} we plot the exponent $\alpha$ and intercept $C$ extracted from the fit as the transition line is crossed by varying interaction strength at fixed density $n=2$.  Plots of this sort enable a simpler and more accurate location of the transition line than an analysis of $C_{SS}(\tau)$ and were used to compute the phase boundary shown in Fig.~\ref{phasediagram}.

\begin{figure}[t]
\begin{center}
\includegraphics[angle=-90, width=0.9\columnwidth]{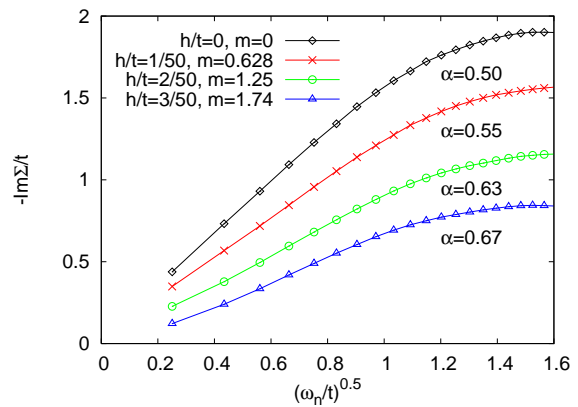}
\caption{Self energy for the majority spin as a function of magnetic field at $U/t=7$, $\beta t=50$ and $n=2$, slightly above the critical point for the glass transition. 
}
\label{field}
\end{center}
\end{figure}

Figures~\ref{qp_doping} and \ref{self} indicate that an approximately square root behavior of the self energy persists for a range of carrier concentrations and interaction strengths near the critical line. 
To further investigate the  effect of frozen spins on $\Sigma$ we have studied the magnetic field dependence of the self energy. As the spins are polarized by the field, the intercept decreases and the apparent power law changes towards the Fermi liquid exponent of $\alpha=1$. However, near the critical point (see Fig.~\ref{field}) $\alpha$ changes slowly and the approximate square root frequency dependence remains visible even for relative magnetizations of the order of $50\%$. 

We speculate that the  $\sqrt{\omega}$ self energy  found here is related to the still-mysterious optical conductivity $\sigma (\Omega)$ of SrRuO$_3$ and CaRuO$_3$  \cite{Kostic98,Dodge02,Lee02}. These materials are pseudocubic perovskites with two holes in the $t_{2g}$ $d$ shell and their behavior should be roughly modelled by our $3$-orbital calculation at carrier density $n=2$ (we neglect effects such as the buckling of octahedra or spin-orbit coupling).  In these materials $\sigma(\Omega)$ varies approximately as $1/\sqrt{\Omega}$ at higher frequencies,  but rolls over (in a way which depends on temperature and on the material) to an approximately constant behavior.   In the momentum-independent self energy approximation, vertex corrections may be neglected and if the self energy is small compared to the bare electron bandwidth the real part of the optical conductivity may be written  ($f$ is the Fermi function)
\begin{align}
\sigma(\Omega)&\approx\frac{1}{\Omega} \int \frac{d\omega}{\pi} \frac{f(\omega+\Omega)-f(\omega)}{\Omega-\Sigma(\omega+\Omega)+\Sigma^{*}(\omega)}
\label{cond}
\end{align}
so that a square-root self energy implies that the conductivity varies as $1/\sqrt{\Omega}$ for frequencies such that $\Sigma(\Omega)>\Omega$. Our finding, that the phase boundary  runs close to $n=2$ for a range of interaction strengths suggests that an extreme degree of fine tuning is not necessary to produce a square root self energy. It is important to note that in SrRuO$_3$ the square root behavior persists well into the ferromagnetically ordered phase at low temperature whereas in CaRuO$_3$ the conductivity develops a nonmonotonic structure at low $\Omega$ and higher $T$. 

The spin-freezing transition itself is of theoretical interest. 
Studies, to be presented elsewhere, of the model without the DMFT self consistency condition show that the transition exists at the level of the impurity model. 
We have also studied the degree to which the different eigenstates of $H_\text{loc}$ are represented in the partition function. We find that for $J>0$  at couplings  ($U \gtrsim 4t$) only a few states are relevant.  The large $U$-density driven transition is marked by a change in the dominant states from the one-electron states $S=1/2$, $L=1$ to a nine-fold degenerate manifold of  two electron states with $S=1$ and $L=1$, with the two manifolds becoming degenerate at the transition.  The interaction-driven transition is on the other hand marked by a change in the weight of the two subleading states $S=1/2$, $L=1$ and $S=3/2$ $L=0$, implying a change in the magnitudes of  coupling strengths.  
The  orbital degree of freedom is important to stabilize the metallic phase at relevant interaction strengths (the two orbital model with two electrons and $J/U=1/6$ is insulating for $U\gtrsim 3.7t$ \cite{Werner07Crystal}). Trying to suppress the $L=1$ orbital angular momentum states by applying a crystal field rapidly leads to an insulator.

In conclusion, we have shown that in a model, relevant to transition metal oxides with partly filled $d$-shells,  with several electrons in a three-fold degenerate level, an apparent spin freezing transition occurs. While it is possible that the effects could be due to a rapid decrease of the spin coherence scale to values  below the range accessible to us, the square root self energy and $T$-linear spin-spin correlation function are strong evidence for an actual $T=0$ transition. The frozen moment phase results from a calculation in which spin rotation symmetry was enforced and may thus be preempted by a state with broken spin rotation symmetry. The exact nature of the broken symmetry state will depend upon the underlying lattice (for the models studied here neither ferro nor antiferromagnetic solutions are stable; we expect that the true ground state is either an incommensurate magnet or is phase separated). However, we emphasize that in contrast to the situation in one-band models, a transition remains even if long ranged order is suppressed in the calculation. 

Our results open many new research directions.  The theory of the multielectron Mott transition must be reexamined, as our results suggest that even for models with high degrees of spin and orbital symmetry a paramagnetic metal-paramagnetic insulator transition generically does not exist (at least for more than one electron per site). The magnetic phase diagram of multiorbital models must be explored. We have shown that this new transition is driven by an apparently new class of impurity-model phase transitions which should be classified and explored. The interplay with lattice-driven crystal field splittings and the orbitally selective Mott transition \cite{Costi07,Werner07Crystal} is an open and interesting question.   More generally the implications of the rich structure of the Slater-Kanamori interactions for the properties of the strongly correlated metal phase deserve further investigation.

\acknowledgements

The calculations have been performed on the Hreidar beowulf cluster at ETH Z\"urich, using the ALPS-library \cite{ALPS}. PW and AJM acknowledge support from DOE-ER-16469 and EG and MT from the Swiss National Science foundation.

\end{document}